\documentclass{aa}

\newcommand{\C}{\v{C}erenkov }
\newcommand{\etal}{{et al. }}

\begin{document}

\thesaurus{08(09.19.2; 09.09.1 W28; 03.13.2; 13.07.2)}

\title{Observations of the supernova remnant W28 at TeV energies}

\author{G.P. Rowell\inst{1}\thanks{\emph{Present address:} Max Planck Institut f\"{u}r Kernphysik, Heidelberg D-69029, Germany}
        \and T.~Naito\inst{6} \and S.A.~Dazeley\inst{2} \and P.G.~Edwards\inst{3} \and 
        S.~Gunji\inst{4} \and T.~Hara\inst{6} \and J.~Holder\inst{1} \and A.~Kawachi\inst{1}
        \and T.~Kifune\inst{1} \and Y.~Matsubara\inst{8} \and Y.~Mizumoto\inst{9} \and 
        M.~Mori\inst{1} \and H.~Muraishi\inst{10} \and Y.~Muraki\inst{8} \and
        K.~Nishijima\inst{7} \and S. Ogio\inst{5} \and J.R.~Patterson\inst{2} \and 
        M.D.~Roberts\inst{1} \and T.~Sako\inst{8} \and K.~Sakurazawa\inst{5} \and
        R.~Susukita\inst{11} \and T.~Tamura\inst{12} \and T.~Tanimori\inst{5} \and 
        G.J.~Thornton\inst{2} \and S.~Yanagita\inst{10} \and T.~Yoshida\inst{10} \and 
        T.~Yoshikoshi\inst{1}}

\mail{Gavin.Rowell@mpi-hd.mpg.de}

\institute{Institute for Cosmic Ray Research, University of Tokyo, Tokyo~188-8502, Japan \and
           Dept. of Physics and Math. Physics, University of Adelaide~5005, Australia \and
           Institute of Space and Astronautical Science, Kanagawa 229-8510, Japan \and
           Dept. of Physics, Yamagata University, Yamagata~990-8560, Japan \and
           Dept. of Physics, Tokyo Institute of Technology, Tokyo~152-8551, Japan \and
           Faculty of Management Information, Yamanashi Gakuin University, Yamanashi~400-8575, Japan \and
           Dept. of Physics, Tokai University, Kanagawa~259-1292, Japan \and
           Solar-Terrestrial Environment Lab., Nagoya University, Aichi~464-8601, Japan \and
           National Astronomical Observatory of Japan, Tokyo~181-8588, Japan \and
           Faculty of Science, Ibaraki University, Ibaraki 310-8512, Japan \and
           Institute of Physical and Chemical Research, Saitama 351-0198, Japan \and
           Faculty of Engineering, Kanagawa University, Kanagawa~221-8686, Japan}

\date{Received 19 October, 1999; accepted 1 April, 1998}  

\titlerunning{Observations of the SNR W28 at TeV energies}
\authorrunning{G.P. Rowell et al.}

\maketitle

\begin{abstract}

The atmospheric \C imaging technique has been used to search for point-like
and diffuse TeV gamma-ray emission from the southern supernova
remnant, W28, and surrounding region. The search, made with 
the CANGAROO 3.8\,m telescope,
encompasses a number of interesting features, 
the supernova remnant itself, the EGRET source 3EG J1800$-$2338,
the pulsar PSR~J1801$-$23, strong 1720 MHz OH masers and molecular clouds on the north and east
boundaries of the remnant.
An analysis tailored to extended and off-axis point sources was used, and
no evidence for TeV gamma-ray emission from any of the features described above was found
in data taken over the 1994 and 1995 seasons. Our upper limit
($E>1.5$ TeV) for a diffuse source of radius 0.25$^\circ$ encompassing both molecular clouds 
was calculated at 6.64$\times10^{-12}$ cm$^{-2}$s$^{-1}$ (from 1994 data), and interpreted within
the framework of a model predicting TeV gamma-rays from shocked-accelerated hadrons.
Our upper limit suggests the need for some cutoff in the parent spectrum of accelerated hadrons and/or 
slightly steeper parent spectra than that
used here ($-$2.1). As to the nature of 3EG J1800$-$2338,  it possibly does not result entirely from $\pi^\circ$ decay,
a conclusion also consistent with its location in relation to W28.

  \keywords{Gamma rays: observations - supernova remnants: individual: W28}

\end{abstract}

\section{Introduction}
 \label{sec:intro}

Supernova remnants (SNRs) have long been thought to be the dominant source of
cosmic rays (CR) at energies below 100 TeV (for a review see e.g.
Blandford \& Eichler \cite{Blandford:1}).  SNR, via the diffusive shock
process, are able to accelerate electrons and hadrons and meet the
energetics of the observed cosmic rays. The TeV gamma-ray flux
predicted from SNR is the most accessible tracer of CR acceleration
and its detection would be convincing evidence for the SNR origin of
galactic CR. Models of the TeV gamma-ray emission from SNR 
predict distinct spectral features, according to the hadronic and/or
electronic nature of the parent CR accounting for the gamma-ray
flux (see Drury \etal \cite{Drury:1}, Naito and Takahara \cite{Naito:1},
Baring \etal \cite{Baring:1}, and references therein for a summary).

Ground-based surveys of SNR at gamma-ray energies (TeV to PeV) have been carried out on 
several promising northern hemisphere candidates (e.g. IC443, Tycho's SNR, W51, W44, G78.2$+$2.1). The Whipple  
(Buckley \etal \cite{Buckley:1}, Lessard \etal \cite{Lessard:1}), HEGRA 
(Hess \etal \cite{Hess:1} at TeV energies, and Prosch \etal \cite{Prosch:1} at multi-TeV energies), CAT (Goret \etal \cite{Goret:1}) and CYGNUS (Allen \etal \cite{Allen:1}) groups have 
reported upper limits. Recently however, the HEGRA has seen marginal evidence for TeV gamma-rays from the young SNR Cas-A, after deep
observation (P\"{u}hlhofer \etal \cite{Puhlhofer:1}). In the southern hemisphere, the CANGAROO has reported the 
detection of TeV gamma-rays from SNR SN1006 (Tanimori \etal \cite{Tanimori:1}) and SNR RX J1713.7$-$3946
(Muraishi \etal \cite{Muraishi:1}), and if confirmed,
will be strong evidence in favour of the production of cosmic rays electrons in SNR. 
  
W28 (also SNR G6.4$-$0.1 from Green \cite{Green:1})
is considered an archetypal composite (mixed or M-type) supernova
remnant, characterised by a centrally filled X-ray and shell-like radio
morphology (Rho \& Petre \cite{Rho:1}, Long \etal \cite{Long:1}).  The
ROSAT X-ray emission appears best explained by a thermal model (Rho
\etal \cite{Rho:2}) although Tomida \etal (\cite{Tomida:1}) from the analysis of ASCA data, has suggested the
presence of a weak a non-thermal component in the south west region.
The limb-brightened radio emission (20, 6 \& 2cm) shows a synchrotron spectrum of varying spectral index
(Andrews \etal \cite{Andrews:1}). A radio point source at $l=6.6^\circ, b=-0.16^\circ$ (G6.6$-$0.1) is defined (Altenhoff \etal \cite {Altenhoff:1}, Andrews \etal \cite{Andrews:1}), hereafter referred as A83 in this paper.
A glitching radio pulsar, 
PSR~J1801$-$23 (PSR~B1758$-23$, $P=$416ms, $\dot{P}=113\times 10^{-15}$ ss$^{-1}$), lies at the northern 
radio edge (Kaspi \etal \cite{Kaspi:1}). 
An upper limit to this pulsar's characteristic age is estimated at 58\,000 years,
and it's spin-down luminosity ($\dot{E} \sim 6.2\times 10^{34}$ erg s$^{-1}$) is at the lower 
edge of luminosity values when compared to the known gamma-ray (EGRET \& COMPTEL) pulsars. 

The age of W28 is estimated (Kaspi \etal \cite{Kaspi:1}) in the range
35\,000 to 150\,000 years, with upper and lower limits taken from
the assumptions that W28 is currently in either the radiative or Sedov
phases of expansion. 
According to Kaspi \etal \cite{Kaspi:1}, the distance of PSR~J1801$-$23 (9 to 16.5 kpc) derived from it's dispersion
measure (DM) appears inconsistent with 
that derived for the remnant. Estimates for the remnant distance are set at 1.8 kpc (Goudis \cite{Goudis:1}
$\Sigma$-D relation) and 3.3 kpc respectively (Lozinskaya \cite{Lozinskaya:1}, from mean optical velocities), 
indicating that the pulsar/W28 association is possibly a
line-of-sight coincidence. However, Frail \etal (\cite{Frail:1}) have noted the large uncertainty in using
the DM as a distance estimate for this pulsar due to the high concentration of ionised material in the 
line of sight, and conclude there is sufficient evidence for the pulsar/remnant association. 
The unidentified
EGRET source 3EG J1800$-$2338 (95\% error circle 0.32$^\circ$ radius)
(Hartman \etal \cite{Hartman:1}), listed as 2EG J1801$-$2312 in the second EGRET
catalogue (Thompson \etal \cite{Thompson:1}),
lies on the edge of the radio shell and was thought to be associated with the remnant (Esposito
\etal \cite{Esposito:1}, Zhang \& Cheng \cite{Zhang:1}). 3EG J1800$-$2338 has a relatively hard spectral
index (Hartman \etal \cite{Hartman:1}) 
with no apparent sign of a turnover at 1\,GeV (Merck \etal \cite{Merck:1}).
Lamb \& Macomb (\cite{Lamb:1}) also point out that 3EG J1800$-$2338
is visible above 1 GeV at 5.4$\sigma$ significance, and is centred very close to the
A83 radio position.
The 3EG position of the EGRET source is displaced by
about 0.5$^\circ$ relative to the 2EG position, yet still lies comfortably within the SNR
radio shell, and remains a strong example of an EGRET source/SNR association (Romero \etal
\cite{Romero:1}). The 3EG error circle however, now excludes PSR~J1801$-$23 and the molecular clouds. 

\begin{figure}
   \vspace{8cm}
   \includegraphics{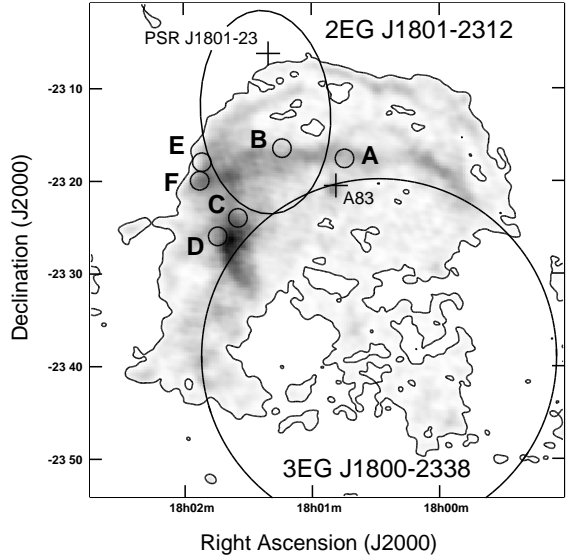}
   \caption{Radio continuum (327 MHz) for W28,
            adapted from Claussen \etal (\protect\cite{Claussen:1}). 
          Included are the positions (and error circles) of the EGRET source 3EG\,J1800$-$2238
          (2EG\,J1801$-$2312 in the 2nd EGRET catalogue), OH maser sites A to F, PSR\,J1801$-$23 
          (also the tracking position for 1994 data) and the radio point source, labelled A83 
          (Andrews \etal \cite{Andrews:1}), is the tracking position for 1995 data. 
          See Sect.~\ref{sec:res} for a discussion.}
   \label{fig:w28_claussen}
  \end{figure}

W28 lies in a complex region of the galactic plane with many HII
regions and dense molecular clouds (Wootten \cite{Wootten:1}) contributing to the 
ISM surrounding the SNR. Over forty OH (1720\,MHz) maser
emission sites are concentrated at the eastern and northern edges of the SNR (Claussen \etal
\cite{Claussen:1}), along the SNR and molecular cloud interface. 
The distribution of shocked and unshocked gas in this region is also consistent with the
idea of the SNR shock passing through the cloud (Arikawa \etal \cite{Arikawa:1}).
OH maser emission (1720 MHz) is considered a strong indicator of collisional pumping
with matter densities $\sim10^5$ cm$^{-3}$ (Claussen \etal \cite{Claussen:1,Claussen:2}).
Enhanced levels of TeV $\gamma$-ray emission via the decay of neutral pions 
may be expected from such areas associated with the masers and molecular cloud 
(Aharonian \etal \cite{Aharonian:1}). Fig. ~\ref{fig:w28_claussen}
indicates the sites of interest in relation to the radio continuum emission (327 MHz).
The presence of these interesting objects make W28 a prime southern hemisphere candidate for 
study at TeV gamma-ray energies.

We report here on the comprehensive analysis of data taken in 1994 and
1995 with the CANGAROO 3.8 metre telescope.  This work follows analysis
of data taken in 1992 (Kifune \etal \cite{Kifune:1}) in which weak
evidence for a gamma-ray signal was reported. At that time, only ON
source data were collected, making an estimation of the background
rate difficult.  Mori \etal (\cite{Mori:1}) reported briefly on
an analysis of 1994 data centred on PSR~J1801$-23$, in
which a $\pm0.7^\circ$ field of view was searched. Both ON and OFF
source data were collected and no evidence for TeV $\gamma$-ray
emission was seen from various point-like sources including 
the pulsar and both radio and X-ray maxima. The 1995 data were centred
on the radio position A83, located $\sim0.3^\circ$ away from PSR~J1801$-23$. 
A search for point-like and diffuse sources of TeV
emission was carried out on the 1994 and 1995 datasets out to $\pm
1^\circ$ from the tracking centre of each dataset, using an
extended source analysis. We have used an
improved set of cuts to those used in the analysis of data taken on
the Vela Pulsar/Nebula (Yoshikoshi \etal \cite{Yoshikoshi:1}). These cuts
were designed to minimise the loss of gamma-ray sensitivity for off-axis sources and
in particular maintain reliable statistics over the search region.

 \section{The CANGAROO 3.8m telescope}

The CANGAROO 3.8\,m imaging telescope is located near Woomera,
Australia ($137^\circ 47^\prime$E, $31^\circ 06^\prime$S, 160m
a.s.l.). The 3.8 metre diameter mirror, of focal length 3.8 metres, is
used to image the \C emission from gamma-ray and cosmic-ray induced
extensive air showers (EAS) onto a high resolution multi-phototube
camera.  The camera consisted of 224 photomultiplier tubes in 1994.
In April 1995 an extra 32 tubes were added to the corners of the
camera bringing the total to 256 tubes in a 16$\times$16 square grid
arrangement. However the extra corner tubes have not been used in the
analysis of the 1995 W28 dataset in order to retain consistent imaging
properties with the 1994 dataset.  Each camera tube is a Hamamatsu
R2248 with a photocathode size 0.12$^\circ \times 0.12^\circ$ on a
side, and the total field of view of the camera is 2.9$^\circ$. An
event trigger is registered when the summed output of triggered tubes exceeds
a preset threshold, denoted HSUM. Images with a minimum of
between 3$\sim$5 tubes, depending on the image's 
compactness, trigger the telescope.
A vertical event rate due to
cosmic rays of $\sim$2 Hz is achieved. Monte Carlo simulations of the
telescope indicate an gamma-ray energy threshold of $\sim$1.5 TeV
at the vertical (Roberts \etal \cite{Roberts:1}), where the energy threshold
is defined as that representing the half-maximum of the differential distribution of triggered energies.
Tracking calibration is performed by monitoring the paths of bright (visual
mag. 3--6) stars in the field of view, providing an absolute tracking
accuracy of $\sim 0.02^\circ$ (Yoshikoshi \cite{Yoshikoshi:2}).
A more detailed technical description of the telescope appears in Hara \etal
(\cite{Hara:1}).

Data are recorded on clear moonless nights. An ON source run is
generally followed by an OFF source run displaced in right ascension
to provide a background run of matching zenith and azimuth angle
distributions. However, since small sections of data from both
observation seasons were removed due to cloud effects, a
normalisation, described later, was used in estimating the
statistical significance of any ON source excess. Pulse charge (ADC) and timing
(TDC) information for each tube is recorded for each event. 
Calibration of the ADCs and TDCs is achieved by recording events
triggered with a blue LED flasher before
each observation. Tube signals are accepted as part of an image if
they meet a number of criteria:
  \begin{enumerate}
   \item The TDC value of a tube must lie between $\pm 37.5$ns, referenced
         against the event trigger time. The event trigger time is
         registered when the HSUM threshold is met.
   \item The ADC value must be greater than one standard deviation above the RMS 
         noise (comprising skynoise and electronic noise) for that tube. 
   \item The tube must not be isolated. An isolated tube is one which is not 
         adjacent to any other accepted tube.
   \item The tube must not have an outlying relative gain value. Tubes with relative gains outside
         the range 1.0$\pm$0.3 contribute significantly to trigger differences 
         across the camera face. This factor is 
         particularly important when comparing regions over the entire camera face. 
  \end{enumerate}

The telescope is an altitude-azimuth type, which
introduces a rotation of the camera relative to the sky about the tracking position 
during data collection. A `de-rotation' is applied to the tube positions to account for this effect, 
and is necessary when considering off-axis sources.
The images are parameterised according to the moment-based method of
Hillas (\cite{Hillas:1}). 

Pre-processing steps designed to minimise the effects of electronic interference
are described below. The camera is divided into groups of eight tubes 
which share common high voltage and other circuitry, and a
special cut, {\em box}
(Yoshikoshi \cite{Yoshikoshi:2}), is designed to remove images
arising from electronic contamination, and are concentrated in only one or two tube boxes, 
This {\em box} cut, in combination with a total ADC sum 
({\em adc}) cut rejecting events with fewer than 200
ADC counts, is very effective at removing such artifacts.
Monte Carlo simulations show that the {\em
box} cut does not reduce the power of the image cuts, and that the
optimum {\em adc} cut lies at $\sim$200 ADC counts.  

Mirror degradation resulted in an event rate drop by about a factor of two
from 1994 to 1995, indicating a higher energy threshold for the 1995
dataset. The results quoted in this work are
normalised to a 1.5 TeV threshold for gamma-rays, using different
raw triggering efficiencies for gamma-rays (Sect.~\ref{sec:extend}), which take into account the
increase in trigger threshold between 1994 and 1995. In addition,
only events with {\em width} $>=$ 0.01 and with the number of triggered tubes,
{\em ntubes}$\geq 4$ are accepted.
The cuts described above are termed {\em noise}
cuts. 

The ON--OFF statistical significance is calculated following Li \& Ma
(\cite{Li:1}), before and/or after application of all image cuts.
  \begin{equation}
   S = \frac{ON - \beta \, OFF}{\sqrt{ON + \beta^2 \,OFF}}
   \label{eq:beta}
  \end{equation} 
and is used to assess the likelihood of a gamma-ray signal.
In order to account for the mismatch of observation times
between ON and OFF source data (and hence zenith angle-dependent event
rates), and trigger rate differences due to subtle changes in weather
conditions and/or telescope response during observation runs,
a normalisation is applied to the ON--OFF statistical significance.
This normalisation, $\beta$ is defined as the ratio of the events available for
image parametrisation, i.e. after {\em noise} cuts. A final systematic
check on the ON--OFF statistics after application of all cuts,
performed on a run-by-run basis, is explained in
Sect.~\ref{sec:res}.

 \section{Extended source analysis}
   \label{sec:extend}
The analysis of extended sources has required the development of new
techniques in TeV gamma-ray astronomy (Akerlof \etal \cite{Akerlof:1},
Hess \etal \cite{Hess:1}, Buckley \etal \cite{Buckley:1}, Connaugton
\etal \cite{Connaugton:1}). These methods use
shape parameters such as {\em width} and {\em length} in combination
with source position-dependent orientation and location cuts (e.g. {\em
alpha, asymmetry} and {\em dis}).  The orientation and location
parameters are recalculated at every trial source position and a
skymap is created. In the analysis of CANGAROO Vela Pulsar data,
Yoshikoshi et al.\ (\cite{Yoshikoshi:1}) used a set of cuts based on {\em
alpha, length, width, concentration} and {\em dis}. A continuous
probability distribution was initially used to locate the position of
the most significant point in the skymap.  A gamma-ray flux at this
point was then estimated from the ON--OFF excess obtained after using
a combination of shape and location cuts with {\em alpha}$\leq
10^\circ$.  These cuts were optimised using Monte Carlo simulations
and therefore are {\em a priori} decisions. We adopt the same {\em a
priori} philosophy here in order to determine the significance
of any result without the need to consider statistical penalties.  
The set of cuts described here were also used in the
analysis of Centaurus~A data described by Rowell \etal (\cite{Rowell:1}).
The Monte Carlo simulation package, MOCCA92 (Hillas \cite{Hillas:3}), 
was used to generate \C images
from extensive air showers (EAS). Gamma-ray primaries were sampled
from a power law above 0.8\,TeV with integral spectral index $-$1.6. 
Cosmic ray primaries, represented by a combination
of proton, helium and nitrogen primaries, were sampled from a power law
of spectral index $-$1.65 above 1.5\,TeV.

\begin{figure*}
   \vspace{13cm}
   \includegraphics{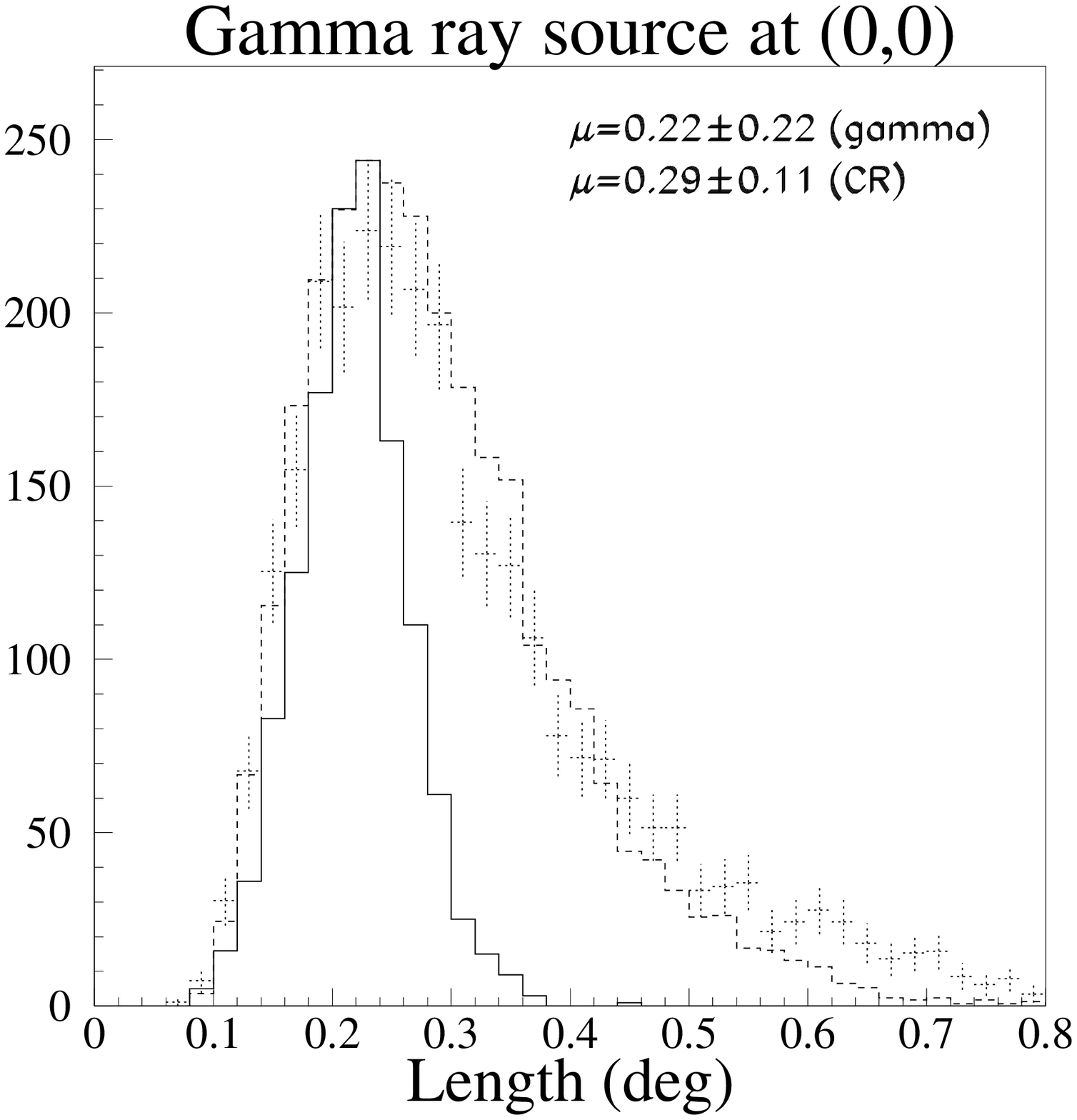}
   \includegraphics{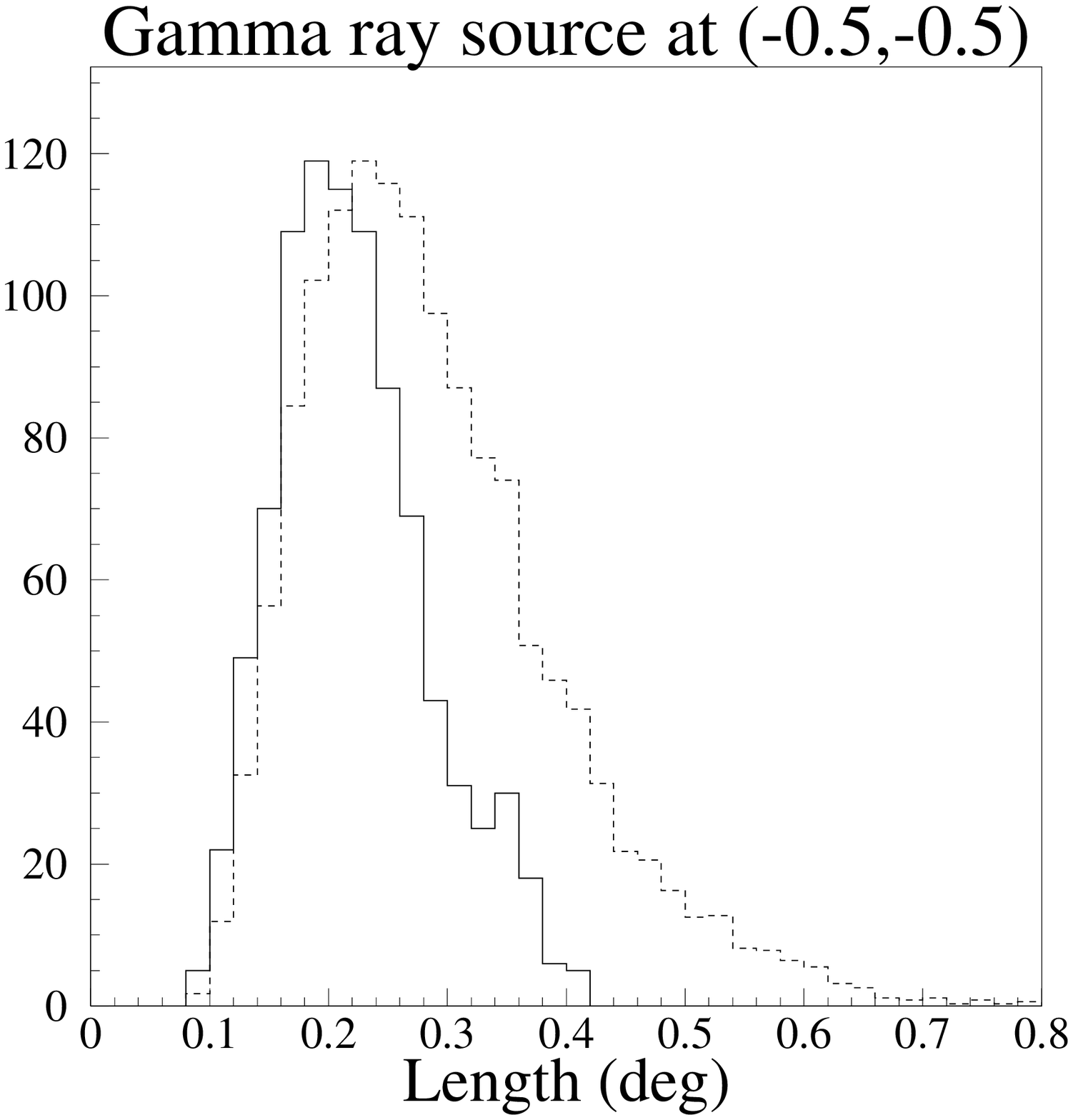}
   \includegraphics{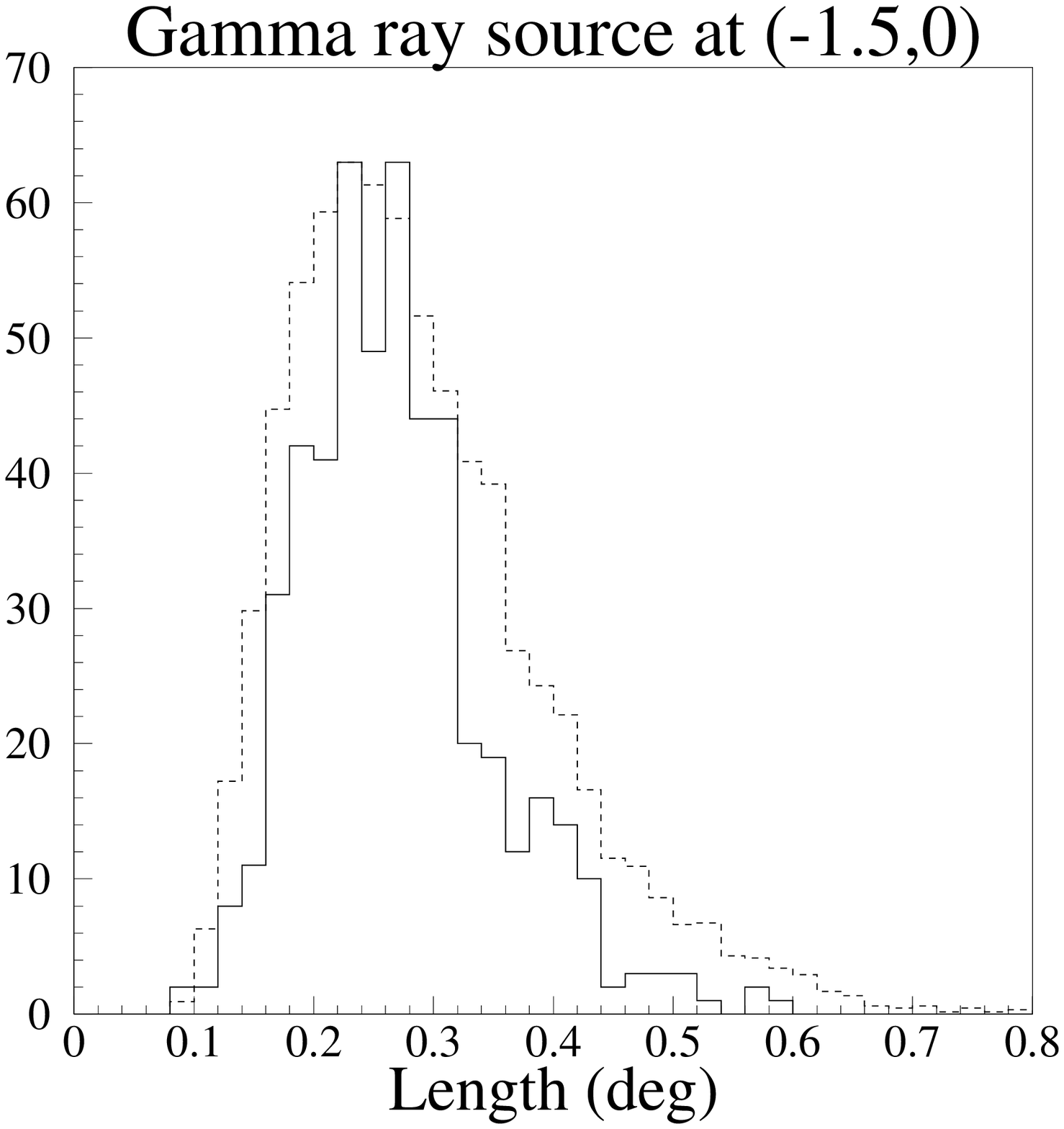}
   \includegraphics{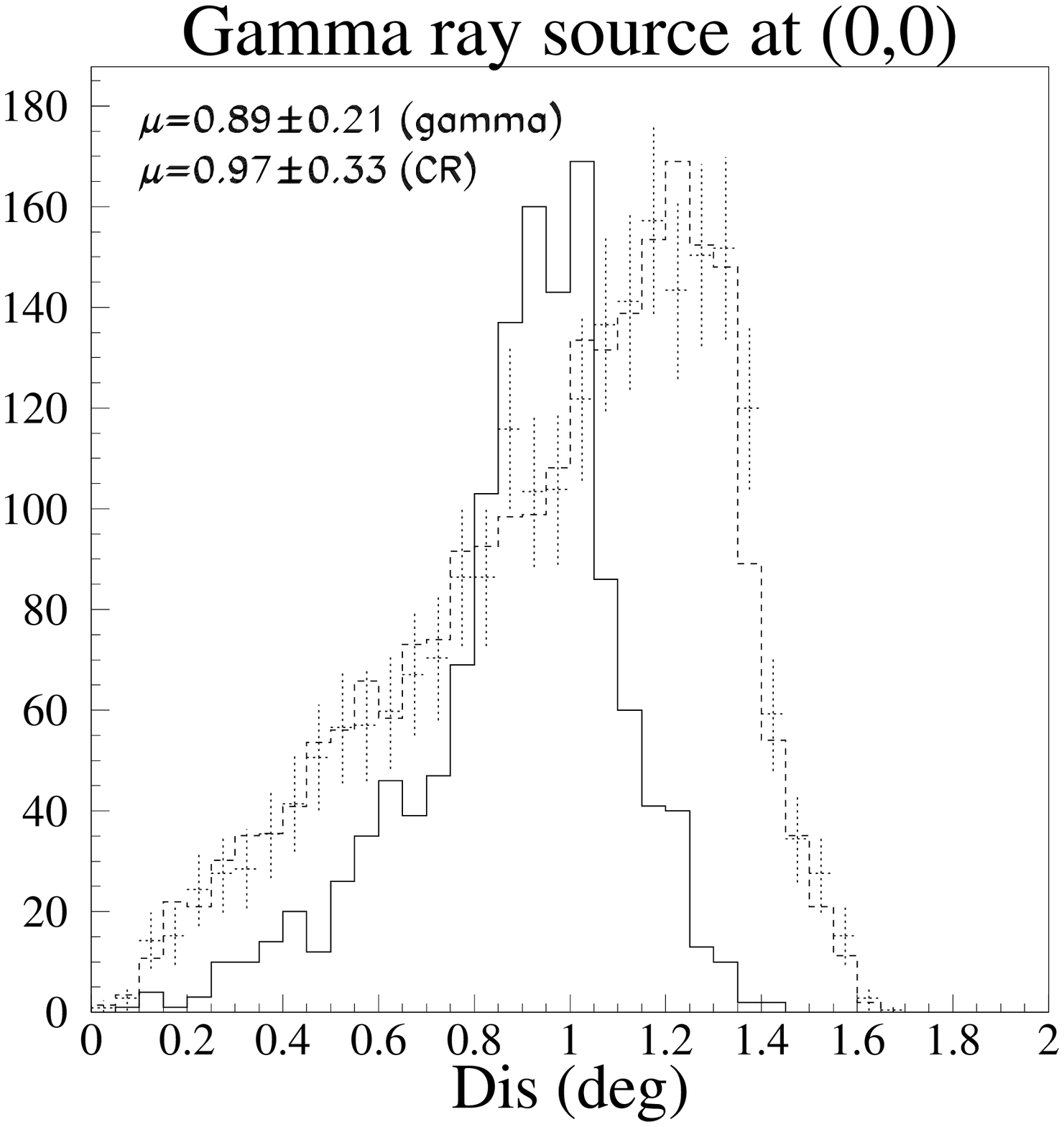}
   \includegraphics{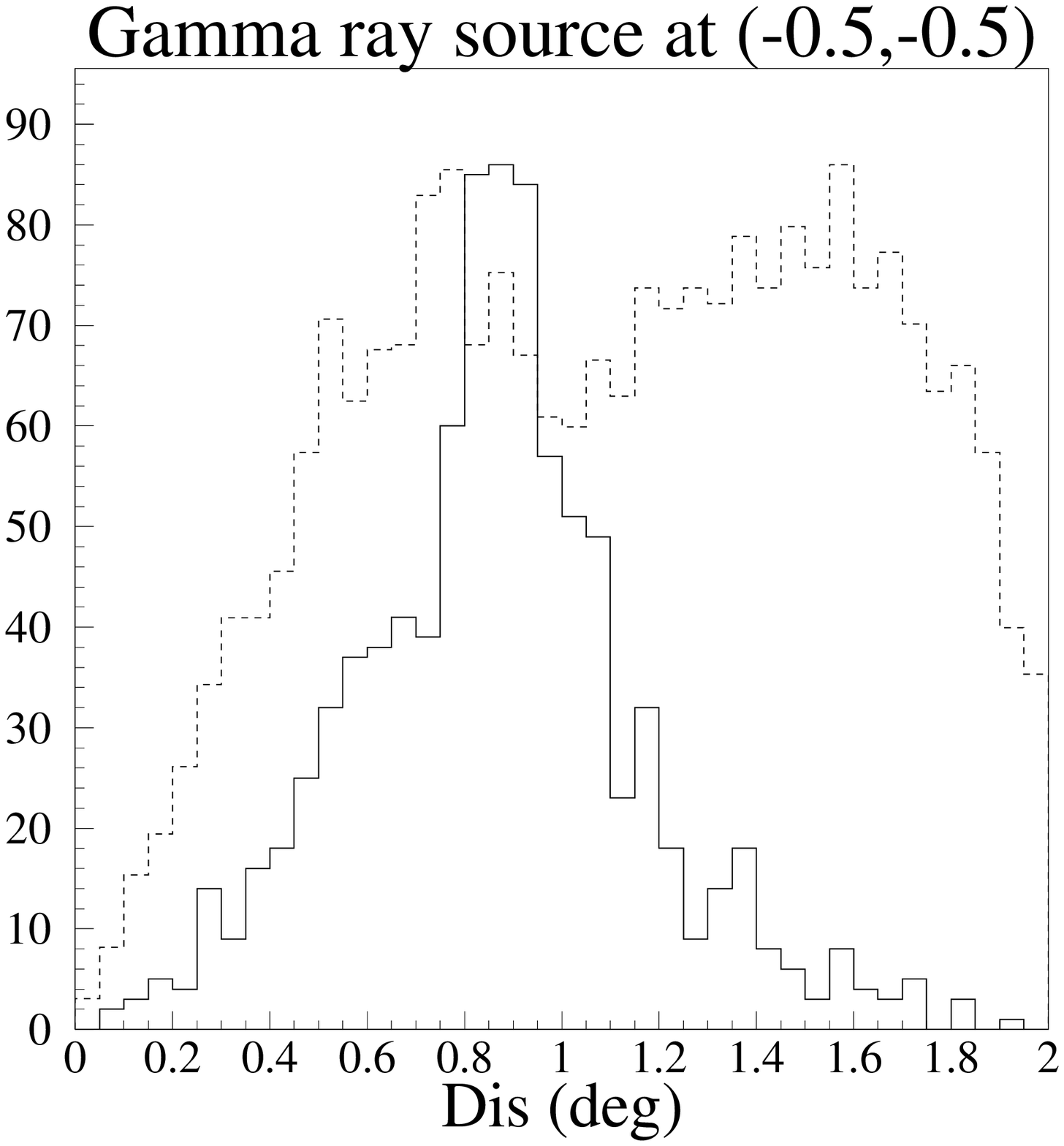}
   \includegraphics{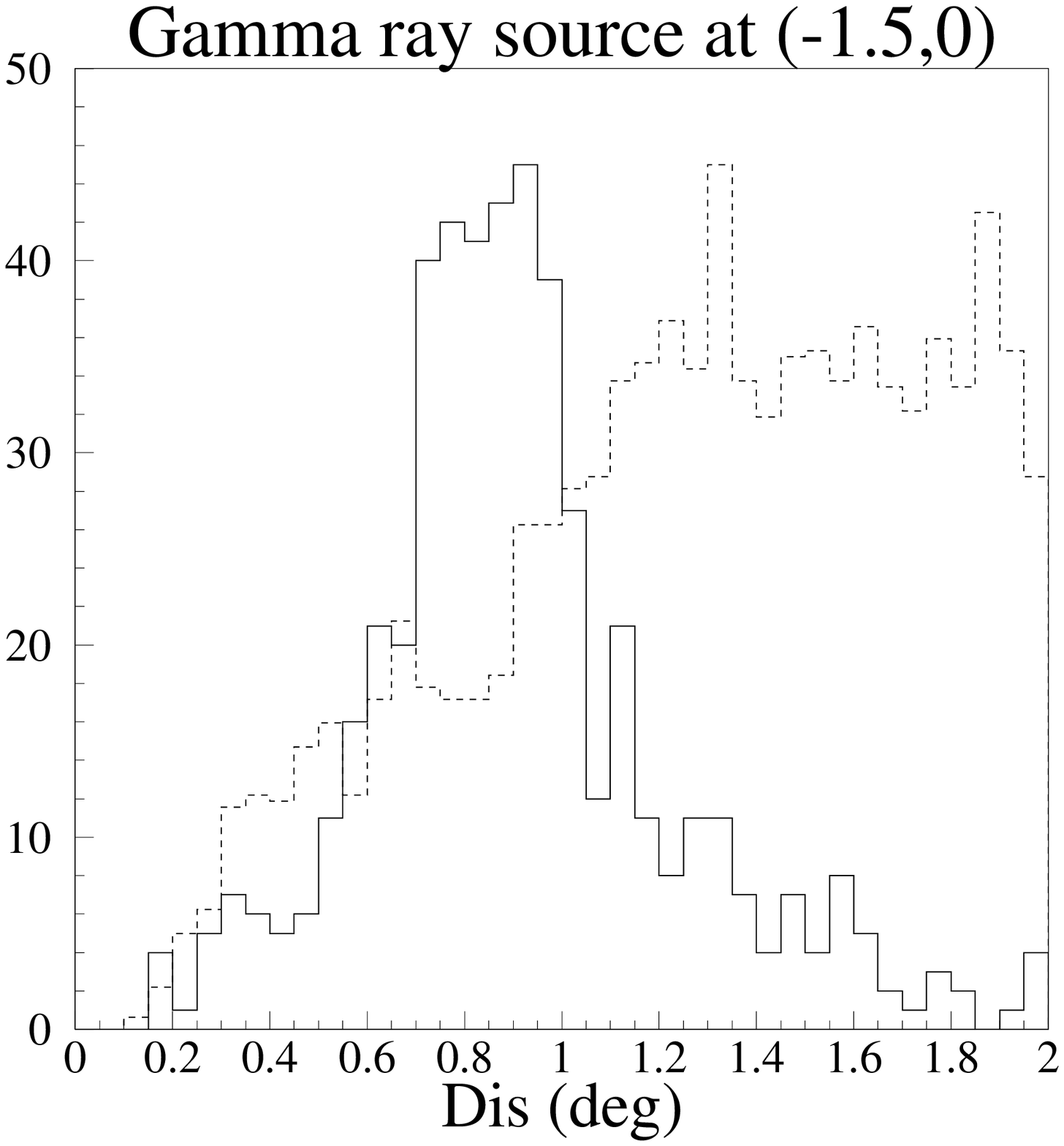}
   \caption{Distributions of {\em length} and {\em dis} for gamma-rays 
            (solid lines) as a function of gamma-ray source radial distance 
            (0.0$^\circ$, 0.71$^\circ$ and 1.5$^\circ$) from the camera centre. 
            Cosmic rays (dashed lines) are included for comparison. The co-ordinate 
            origin of the camera is shifted to the gamma-ray source position 
            when calculating {\em dis}. {\em Noise} cuts have been applied. In the 
            leftmost plots, real OFF source data, normalised to the simulated CR
            events are included (dotted error bars), as well as mean values for the
            simulated gamma and CR.}
   \label{fig:ledis}
  \end{figure*}

In designing an analysis suited to off-axis and extended sources, it
is important to consider the off-axis sensitivity of the CANGAROO
camera, particularly considering that it has a relatively small field
of view and operates at a gamma-ray threshold of about 1.5\,TeV.  The
behaviour of various image parameters as a function of gamma-ray
source position has been investigated. A gamma-ray point source was
placed at six positions across the camera and each position
considered independently. The resulting distributions of {\em width} and {\em
miss} showed little variation with source distance from the camera
centre, in contrast to those of {\em length} and {\em dis} (Fig.~\ref{fig:ledis}).
For increasing source distances from the camera centre, larger values
of {\em length} and {\em dis} are possible, behaviour which is readily
understood in terms of camera edge effect reduction on one side as the
source approaches the camera edge. In particular, the {\em length}
parameter for gamma-rays is very similar to that for cosmic-rays for
sources beyond about 1.0$^\circ$ from the camera centre. Indeed, even
for on-axis sources, the field of view of the 3.8\,m camera imposes
some edge effects, a consequence of operating at a relatively high
gamma-ray energy threshold. Thus, for off-axis sources, a reduced 
gamma ray efficiency will result when using cuts derived for
an on-axis source. 

Firstly, there is a clear case for increasing
the gamma-ray efficiency while maintaining the quality factor on the grounds of improved event 
statistics. An increase in gamma-ray efficiency entails an increase in background
acceptance, thus improving an estimate of the background and diluting
any unaccounted systematic effects. It is also possible to examine the effect
of the gamma-ray efficiency on the ON--OFF significance after cut
application. We can express 
Eq.~(\ref{eq:beta}) in another way, incorporating the gamma-ray efficiency, after 
application of image cuts:
 \begin{equation}
   S = \frac{N_\gamma}{\sqrt{\frac{N_\gamma}{\epsilon_\gamma} + \frac{2N_b}{Q^2}}}
   \label{eq:newsig}
  \end{equation}
where the gamma-ray signal, $N_\gamma$ 
($N_\gamma$=ON--OFF) and the background, $N_b$ ($N_b=$OFF), represent those {\em prior} to image cuts.
The efficiency for gamma-ray selection is given by 
$\epsilon_\gamma$ and that for the background (CR), $\epsilon_b$.
The quality or Q-factor of the cuts is given by $Q=\epsilon_\gamma /\sqrt{\epsilon_b}$ . Here, we set the normalisation 
factor from Eq.~(\ref{eq:beta}), $\beta=1$. We can see that the significance obtained after image cuts is
dependent on the quality factor {\em and}, somewhat slightly, on the gamma-ray cut efficiency.
For sources with a high gamma-ray to CR flux ratio (e.g. $>$0.l) with $Q\sim$4 (as for the CANGAROO telescope), both denominator terms 
are then similar, the sensitivity 
of $S$ to $\epsilon_\gamma$ becomes more apparent. Such a 
situation may arise in the case of searches for bursts from AGN and/or gamma-ray bursts over short 
time scales. For signal to noise ratios expected of SNRs such as W28 however, we would expect only a minor improvement in $S$ 
from the above arguement. Thus, the main motivation for increasing $\epsilon_\gamma$ here is simply to work with increased statistics.
     
\begin{figure}
   \vspace{9cm}
   \includegraphics{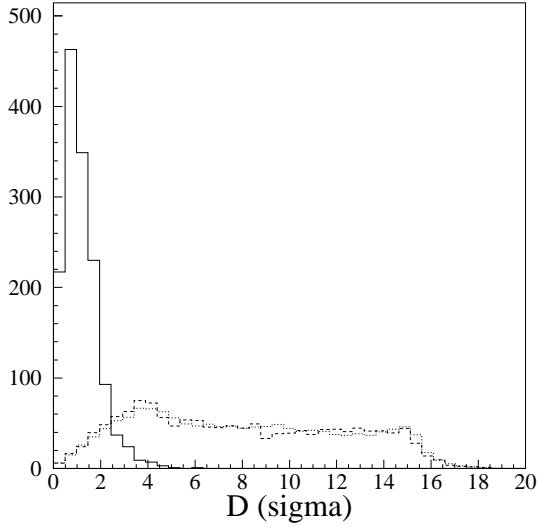}
   \caption{Distributions of $D$ for  simulated gamma-rays 
            (solid line), simulated cosmic rays (dashed line) and real OFF source data (dotted
             line).} 
   \label{fig:D}
  \end{figure}

The variable cuts on {\em length} and for {\em
dis} were incorporated into the cut ensemble.
Along with the 3rd moment of the image, {\em asymmetry},
we use an approximation of the distance between the assumed source
position and calculated source position for each image, $D$. This
distance, expressed in units of standard deviation, is given by:-
  \begin{equation}
   D = \sqrt{ \left( \frac{miss}{\sigma_{miss}} \right)^2 + 
                     \left( \frac{dis-dis_{ex}}{\sigma_{dis}} \right)^2 }
    \label{eq:sdf2}
  \end{equation}
where $\sigma_{miss}, \sigma_{dis}$ are the variances
for {\em miss} and {\em dis}
respectively. The variances, $\sigma_{miss}$ and $\sigma_{dis}$ represent the
transverse and longitudinal errors in the most likely source position
for an image. $D$ can be characterised as the source density function
(SDF).  When used in combination with shape and {\em asymmetry} cuts,
a cut on $D$ provides a gamma-ray acceptance of $\sim$40\% for an
on-axis point source. 
$D$ is similar to the
normalised cluster, or Mahalonobis distance for {\em miss} and {\em
dis} (Hillas \& West \cite{Hillas:2}), although here we are neglecting
cross-term variances. $D$ can also be considered a discrete analogue of the
probability distribution function used by Yoshikoshi \etal (\cite{Yoshikoshi:1}).
The longitudinal error in source location is
greater than the transverse error by about a factor of two. Following
Akerlof \etal (\cite{Akerlof:1}), some reduction in the longitudinal
error can be achieved by making use of the {\em elongation}
(defined as {\em length/width}) of the image such that the expected longitudinal
source distance is given by $dis_{ex} = g (1-(width/length))$ where
$g$ is an empirically derived constant. For the 3.8\,m camera we
derive a value of $g=1.25$ using simulations. Fig.~\ref{fig:D} gives distributions of the
parameter $D$ for simulated gamma-rays and cosmic rays, and a comparison to
real OFF source data.
Linear fits were found for the cut on {\em
length} and value of $\sigma_{dis}$, respectively, as a function of
source displacement from the camera centre.  The total cut combination
is listed in Table~\ref{tab:cutlist}.
  \begin{table}
   \begin{center}
   \begin{tabular}{ll} \hline\hline
    Parameter    & Range \\ \hline
    {\em adc}    & $\ge 200$ ADC counts \\
    {\em box}    & $\le 0.99$ \\
    {\em ntubes} & $\ge 4$ \\
    {\em width}  & 0.01$\leq${\em wid}$\leq$0.11 \\
    {\em length} & $\le 0.32+0.10d$ \\
    {\em asymmetry} & $\ge$0.0 \\
    {\em D}      & $\le 2.0 \sigma$ \\
                 & $\sigma_{dis} = 0.21+0.09d$\\
                 & $\sigma_{miss} = 0.09$ for all $d$ \\
                 & $dis_{ex} = 1.25(1.0-\frac{width}{length})$ \\ \hline\hline
   \end{tabular}
   \end{center}
   \caption{Image cuts used in the extended source analysis. 
            See Eq.(\protect~\ref{eq:sdf2}) for a definition of $D$. 
            The parameter $d$ is the assumed source displacement from the 
            camera centre (in degrees).
            The {\em length} cut and value of $\sigma_{dis}$ are dependent 
            upon the assumed source position, and are derived from linear fits.} 
   \label{tab:cutlist}
  \end{table}
These cuts
(Table~\ref{tab:cutlist}) provide a constant gamma-ray efficiency of
$\sim$40\% (and cosmic-ray efficiency) and quality factor $\sim$4 at the {\em same} value of $D$.
Without the variable {\em length} and {\em dis} criteria, the
gamma-ray acceptance quickly reduces to less than 30\% for sources
outside 0.5$^\circ$ from the camera centre for no significant
improvement in Q-factor.  Table~\ref{tab:simresults} gives the
simulated performance of the cut $D$ for various source positions. A
comparison to the performance obtained by the set of cuts used in the
Vela Pulsar analysis (Yoshikoshi et al.\ \cite{Yoshikoshi:1}) is included.
These cuts were used to obtain a significant excess of gamma-ray-like events from a
region displaced 0.14$^\circ$ from the pulsar position. 
The Vela Pulsar cuts are based on the same noise cuts listed in
Table~\ref{tab:cutlist}, except that fixed values of {\em length} and
{\em dis} are used and a cut on {\em alpha}$\leq 10^\circ$ is
substituted for $D\leq 2.0$. The Vela Pulsar cuts do not provide a
constant gamma-ray acceptance over the search region, and the background (cosmic ray)
acceptance also decreases sharply.

  \begin{table}
  \begin{center}
  \begin{tabular}{ccccc} \hline \hline
   \multicolumn{5}{c}{Cuts used in this work (see Table~\protect\ref{tab:cutlist})} \\ \hline
   Source Location & $Q$ & gammas & cosmic rays & $\epsilon_\gamma$ (\%) \\ \hline
   0.0$^\circ$,0.0$^\circ$           & 4.5      & 868    & 77          & 43 \\
   $-$0.5,0.0        & 4.2      & 708    & 84          & 41 \\
   $-$0.5,$-$0.5     & 3.9      & 591    & 90          & 40 \\
   $-$1.0,0.0        & 4.2      & 522    & 80          & 40 \\
   $-$1.0,$-$1.0     & 4.1      & 375    & 89          & 42 \\
   $-$1.5,0.0        & 4.1      & 370    & 88          & 42 \\ \hline
   \multicolumn{5}{c}{Vela Pulsar Cuts (Yoshikoshi \etal \protect\cite{Yoshikoshi:1})} \\ \hline
   0.0,0.0           & 3.6      & 887    & 125          & 44 \\
   $-$0.5,0.0        & 3.5      & 596    & 85          & 35 \\
   $-$1.0,0.0        & 4.3      & 401    & 45          & 31 \\
   $-$1.5,0.0        & 4.8      & 255    & 31          & 29 \\ \hline \hline
  \end{tabular}
  \end{center}
  \caption{Simulated performance of the image cuts of this work (assessed using the quality factor
           $Q$) compared 
           with cuts used in the Vela Pulsar analysis. Since the 
           statistical errors of $Q$ are $\sim10$\%, the Vela 
           Pulsar cuts are not statistically different from those of
           this work. The Vela Pulsar cuts use the same noise
           cuts as defined in Table\protect~\ref{tab:cutlist}, but fixed 
           values of {\em length} and {\em dis}
           (Yoshikoshi \etal \protect\cite{Yoshikoshi:1}).}
  \label{tab:simresults}
  \end{table}

The resulting flux or upper limit from the search is 
found by dividing the ON--OFF excess, 
$N$ (or 3$\sigma$ upper limit thereof), by the position-dependent 
exposure (for an extended source, averaged over the integration region):
\begin{equation}
 F (> E)=\frac{N}{\epsilon_\gamma \bar{\eta} A T} \;\; {\rm photons\;cm^{-2}\;s^{-1}}
 \label{eq:flux}
\end{equation}
where $\bar{\eta}$ is the {\em raw} trigger efficiency for gamma-rays
averaged over the integration region, $\epsilon_\gamma$ is a constant
gamma-ray selection efficiency for the SDF cut, $A$ is the area over
which gamma-rays are simulated ($A = 1.96\times 10^9$ cm$^2$) and $T$
is the total observation time. We set $\epsilon_\gamma$ to be the
average of the simulated gamma-ray cut efficiencies out to $\pm
1^\circ$, i.e. $\epsilon_\gamma$ = 0.41. An unavoidable decline in 
$\eta$ for gamma-rays will occur as the source position
is displaced further off-axis. Simulations show that $\eta$
decreases to less than half the on-axis value at the corners of
the search.  A linear fit was used to
characterise $\eta$ at all points within
$\pm 1^\circ$ such that $\eta_{d} = \eta_0(1.0 - 0.39d)$ as a function
of the source displacement from the camera centre, $d$. We find 
$\eta_0$(1995 data)$ = 0.12$ at the camera centre above the minimum simulated
energy of 0.8 TeV for the 1995 dataset. The raw gamma-ray trigger
efficiency for 1994 data was estimated by scaling the 1995 value by
the ratio of observed event rates after noise cuts, giving 
$\eta_0$(1994 data)$ = 0.21$. In characterising
$\epsilon_\gamma$ and $\bar{\eta}$, we are assuming that the gamma-ray
flux of an extended source is isotropically distributed.  To calculate
the flux applicable to the energy threshold of 1.5 TeV (Roberts
\etal \cite{Roberts:1}) we used the gamma-ray spectral index of --1.6
adopted in the simulations.  For a point-like search, the flux was
taken as that from the point of interest, using a single value of
$\eta$.  The nature of our gamma-ray selection cuts naturally
incorporates the gamma-ray point spread function (PSF). The cut on $D$
accepts events with derived source positions within an optimal
radius. A search for an extended source therefore does not require any
extra area to account for the PSF in addition to the source area
itself.  In an extended source search, at a suitably high number of
assumed source positions in the region of interest, we sum the events
passing all cuts, taking care not to count an event more than once.
Skymaps of the statistical significance, $S$, of a gamma-ray signal were
generated over a $\pm 1^\circ$ area   
at 0.05$^\circ$ steps. At each
grid point, representing an assumed source position, image parameters
were calculated and the number of events passing the cuts of
Table~\ref{tab:cutlist} cumulatively summed for all data. Since the
resolution of the grid (0.05$^\circ$), is smaller than the effective
acceptance area of the cuts (the cut on $D$ alone is more powerful
than a cut on {\em alpha}$\leq$10$^\circ$), each skymap point value will not
be fully independent of its neighbours.

As a final check on data integrity, the distribution of $S$ obtained
on a run-by-run basis (i.e. run-by-run skymap) was quantitatively
assessed for systematic effects. The most important systematic effect
to consider in this type of analysis is the consistency of the trigger
threshold over the entire camera between ON and OFF source runs. Such
an effect is 
difficult to compensate
for after the data are taken. The Kolmolgorov-Smirnoff (KS) test is used
to examine the likelihood that the distribution of $S$ of the skymap
obtained on a run-by-run basis is derived from a normal
distribution. Over the time scales of a single run ($\leq$5 hours), we
do not expect significant contributions from a steady source of TeV
gamma-rays of the strength expected of a SNR or plerion. A relatively
strong KS probability of 4$\sigma$ was used to reject pairs of data
that appeared severely affected by such systematics. For unmatched
pairs, a well-behaved OFF or ON comparison run was used. A total of 10
hours data were rejected using the KS test, representing three ON/OFF
pairs from the 1994 dataset.

 \section{Results}
  \label{sec:res}
  Table~\ref{tab:data_summary} presents a summary of data accepted 
  for analysis from the 1994 and 1995 observing seasons. 
  \begin{table}
  \begin{center}
  \begin{tabular}{lcccc}\hline \hline
                           & \multicolumn{2}{c}{1994 Data} & \multicolumn{2}{c}{1995 Data} \\ \cline{2-5}
                           & ON    & OFF   & ON     & OFF \\
   Time(hrs)               & 20.3  & 23.5  & 37.5   & 30.0 \\
   Images                  & 59909 & 65483 & 97850  & 76595 \\
   After {\em noise} cuts  & 37386 & 40731 & 39354  & 31237 \\
   \multicolumn{1}{c}{ON, OFF normalisation $\beta$}  & \multicolumn{2}{c}{0.92} & \multicolumn{2}{c}{1.26} \\
   \multicolumn{1}{c}{Raw gamma-ray trigger} & & & & \\
   \multicolumn{1}{c}{efficiency $\eta_0$} & \multicolumn{2}{c}{0.21} & \multicolumn{2}{c}{0.12} \\ \hline \hline
  \end{tabular}
  \end{center}
  \caption{Summary of 1994/1995 W28 CANGAROO data accepted for analysis. 
           The normalisation factor $\beta$ used in adjusting the  
           statistical significance of the post-cut ON--OFF excess, 
           and the raw trigger efficiencies for gamma-rays above 
           0.8 TeV ($\eta_0$) are included.}
   \label{tab:data_summary}
  \end{table}

  \begin{figure*}
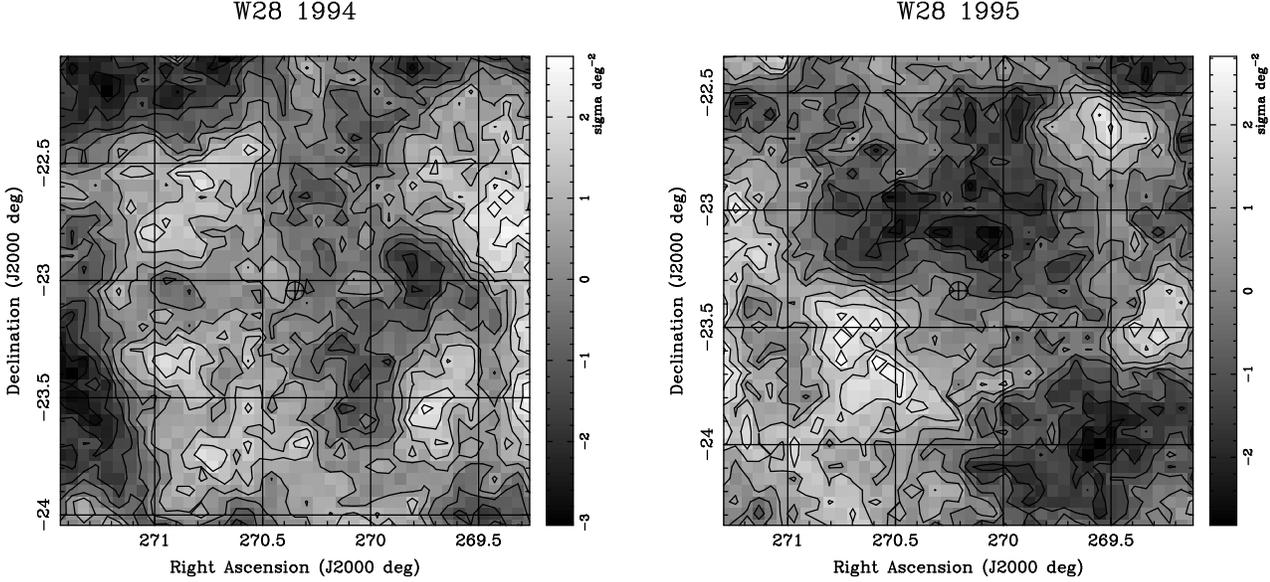


   \vspace{10cm}
   \includegraphics{skymap_w28_1994.ps}
   \includegraphics{skymap_w28_1995.ps}

   \caption{SDF skymaps of ON--OFF significance for W28 1994 data (left), 1995 data
            (right). The skymaps extend $\pm 1^\circ$ from the respective tracking centres
            (indicated by the circled cross) of each year's data. Contour levels indicate
            the same ON--OFF excess over 10 intervals between the maximum and minimum excess values 
            of each skymap.}
   \label{fig:skymaps}

  \end{figure*}
The skymaps of significance $S$
(Fig.~\ref{fig:skymaps}) for both years' data,
do not indicate any significant point-like excesses over a
$\pm 1^\circ$ search for $\gamma$-ray emission.
Table~\ref{tab:sourcelist} summarises the 3$\sigma$ upper limits for the positions of a
number of interesting sites within the W28 region. It should be pointed out that our results will 
have a systematic error of the order $\sim20$\%, based on uncertainties in the trigger conditions, mirror
reflectivity and spectral index adopted in the simulations (Yoshikoshi 
\cite{Yoshikoshi:2}). For example, in the next section we compare our results to a model producing
a gamma-ray flux with spectral index of $\sim -1.1$, harder than the $-1.6$ spectrum used in simulations. Such
a difference however will contribute a systematic of $\leq$5\%.

The following positions were considered as potential sources; the position A83 
given by Andrews \etal (\cite{Andrews:1}), the pulsar PSR~J$1801-23$,
the two strongest (by an order of magnitude) masers, labelled E \& F
by Claussen \etal (\cite{Claussen:1}), and the EGRET source
3EG J1800$-$2338. In addition, an extended region of radius 0.25$^\circ$
centred on a position to encompass the molecular clouds was considered.
An average position was used for the masers E and F since they
are separated by only 0.02$^\circ$.
  \begin{table*}
  \begin{center}
  \begin{tabular}{lcccccccccc} \hline \hline
          &          &         &    \multicolumn{4}{c}{ 1994 DATA }   &    \multicolumn{4}{c}{ 1995 DATA } \\
          & RA       & Dec     &    &     &     & F ($>$1.5 TeV)      &     &     &  & F ($>$1.5 TeV)  \\ 
   Source & (J2000)  & (J2000) & ON & OFF & $S$ & cm$^{-2}$s$^{-1}$  & ON & OFF & $S$ & cm$^{-2}$s$^{-1}$ \\ \hline
   A83$^a$             & 270.21 & $-$23.34 &  644 &  698 &   +0.1 & $<$3.36$\times 10^{-12}$ 
                                           &  632 &  504 & $-$0.1 & $<$2.95$\times 10^{-12}$ \\
   Mol. Clouds$^b$     & 270.35 & $-$23.34 & 2589 & 2843  & $-$0.3 & $<$6.64$\times 10^{-12}$ 
                                           & 2663 & 2163  & $-$0.8 & $<$5.56$\times 10^{-12}$ \\
   PSR J1801$-$23$^c$  & 270.35 & $-$23.04 &  644 &  698 &   +0.1 & $<$3.20$\times 10^{-12}$ 
                                           &  723 &  587 & $-$0.4 & $<$3.32$\times 10^{-12}$ \\
   Masers E \& F$^d$   & 270.47 & $-$23.31 &  765 &  822 &   +0.3 & $<$4.14$\times 10^{-12}$    
                                           &  798 &  643 & $-$0.3 & $<$3.47$\times 10^{-12}$ \\   
   3EG J1800$-$2338$^e$ & 270.12 & $-$23.65&  974 &  965 &   +2.1 & $<$8.82$\times 10^{-12}$ 
                                           &  804 &  575 &   +1.9 & $<$1.18$\times 10^{-11}$ \\
    &   &    & \multicolumn{4}{c}{269.80$^f$, $-$23.59} & \multicolumn{4}{c}{270.43$^f$, $-$23.74} \\ \hline \hline
   \multicolumn{11}{l}{\scriptsize a: Radio point source (Andrews \etal (\protect\cite{Andrews:1})} \\
   \multicolumn{11}{l}{\scriptsize b: Extended source of radius 0.25$^\circ$ encompassing the molecular clouds.} \\
   \multicolumn{11}{l}{\scriptsize c: Point source at pulsar position given by from Kaspi \etal (\protect\cite{Kaspi:1}).}\\
   \multicolumn{11}{l}{\scriptsize d: Point source at average position of 1720 MHz OH maser sites E and F as defined by  
                                       Claussen \etal (\protect\cite{Claussen:1}). See text.}\\
   \multicolumn{11}{l}{\scriptsize e: Highest pointlike significance within 95\% error
                                       circle (0.32$^\circ$ radius) of EGRET source. Quoted position is that for}\\
   \multicolumn{11}{l}{\scriptsize \hspace{3mm} the error circle centre (Hartman \etal \protect\cite{Hartman:1}). See also $f$ below.}\\
   \multicolumn{11}{l}{\scriptsize f: Positions of each pointlike maximum significance within the 3EG error circle.}\\
  \end{tabular}
  \end{center}
  \caption{Details of various sites considered as potential emitters of TeV gamma-rays in the W28 region 
           for CANGAROO 1994 and 1995
           data and corresponding 3$\sigma$ flux upper limits. The statistical significance $S$ is 
           calculated from Eq. (\protect\ref{eq:beta}) with the normalisation factor $\beta$ given
           by Table \protect\ref{tab:data_summary}.}
  \label{tab:sourcelist}
  \end{table*} 
In assuming that the EGRET source was point-like, a search for the
highest point significance within the 95\% error circle was carried
out. Since the statistical degrees of freedom for a non-{\em a priori} search
for point-like emission over the skymap is $\sim$100, 
the highest ON--OFF excess within the EGRET error circle must be interpreted
with a similar statistical penalty in mind.  

The point spread function (PSF) for a pure gamma-ray signal can be
used to assess the location accuracy of the proposed source positions.
Monte Carlo simulations show that the PSF for gamma-rays increases
slightly from a standard deviation of 0.2$^\circ$ on-axis to 0.22$^\circ$ for
sources at the skymap corners. The positions of our candidate
sources (Table~\ref{tab:sourcelist}) lie within 1.0$^\circ$ on-axis.
Taking a conservative estimate of the PSF as 0.22$^\circ$, a
simplistic estimate of the source location error is obtained by
adding in quadrature 0.22/$\sqrt{100}$=0.02$^\circ$ to the the
absolute tracking precision (0.02$^\circ$), giving $\sim
0.03^\circ$. We have used a value of 100 which is representative of
the 3$\sigma$ upper
limit excesses ($N$ in Eq.~(\ref{eq:flux})) calculated here. We can therefore say that the
features listed in Table~\ref{tab:sourcelist} would be resolved by the
CANGAROO 3.8m telescope if they were point sources of TeV gamma
rays.  From Table~\ref{tab:sourcelist} we can also see that
the positions of the highest significance within the EGRET 95\% error
circle for both year's data are also not co-located, with their separation,
$\sim0.6^\circ$, easily exceeding the estimated PSF.

\section{Gamma Ray Production in SNR}
SNR shocks are able to accelerate particles to TeV energies. Gamma-rays are produced
in secondary reactions between these high energy particles and ambient matter and
radiation fields. The decay of $\pi^\circ$, produced in ion-ion collisions, is the  
prime hadronic source of gamma-rays in SNR.
Electrons accelerated to multi-TeV energies in SNR give rise to
bremsstrahlung and inverse Compton (IC) scattering 
$\gamma$-ray production processes in SNR (see e.g. Mastichiadis
\& de Jager \cite{Mastichiadis:1}, Gaisser \etal \cite{Gaisser:1}). The
cosmic microwave background (CMB) is usually considered the dominant soft photon source with
contributions ($\sim$10\%) from the infrared background as seed photons for the IC process. The detailed model of
Baring \etal (\cite{Baring:1}) uses these processes collectively (including synchrotron and bremsstrahlung radiation)
to account for observations from radio to TeV gamma-ray energies for the northern SNR, IC443.

The models of Naito and Takahara (\cite{Naito:1}) and Drury \etal (\cite{Drury:1}) give predictions
of the TeV gamma ray flux due to $\pi^\circ$ decay. As a first attempt at explaining the particle
acceleration processes in W28, we make use of the Naito and Takahara model and compare its predictions of the
$\pi^\circ$ channel
to our results. By considering only the $\pi^\circ$ production channel, a lower limit on the predicted 
TeV gamma flux can be estimated. A proton parent 
spectrum with differential 
index $-$2.1 and exponential cutoff at 100 TeV has been assumed ($dN/dE \propto E^{-2.1} \times \exp(-E/100))$, i.e. consistent with shock acceleration expected in a 
SNR. The predicted TeV flux will scale according to:
\begin{equation}
  F_\gamma \propto\frac{E_{cr} \, n}{d^2}
 \label{eq:fluxscaling}
\end{equation}
where $E_cr$ is the energy available for cosmic ray acceleration, $n$ is the particle number density of the ISM
(cm$^{-3}$) and $d$ is the distance to the SNR (kpc). $E_{cr}$ is some fraction, typically $\sim$10\% of the
total energy of the SNR (canonically 10$^{51}$erg). In fact, the total SNR energy for W28 has been estimated by 
Rho \etal (\cite{Rho:2}) at 4$\times10^{50}$erg from ROSAT X-ray data.

\begin{figure}

   \vspace{8cm}
   \includegraphics{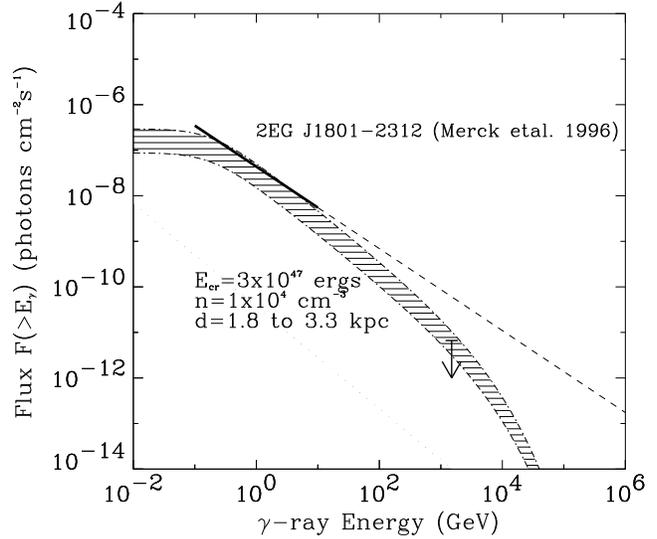}

   \caption{Comparison with our upper limit and the EGRET flux of 
            3EG J1800$-$2338 (Merck \etal 1996) with a model predicting the TeV gamma-ray flux due to 
            $\pi^\circ$ decay (hashed area, Naito and Takahara \protect\cite{Naito:1}). The hashed region 
            is bounded by
            limits on the predicted flux when assuming published ranges of values for various scaling 
            parameters (eq.~\protect\ref{eq:fluxscaling}) defined for the region of shocked gas 
            surrounding W28 (Arikawa \etal \protect\cite{Arikawa:1}). This region is expected to
            dominate those that would produce a TeV gamma-ray flux from the decay of $\pi^\circ$. Our upper
            is for an extended source containing the molecular clouds (see text).
            The model is practically upper-bounded by the flux from 2EG J1801$-$2312, and parent-proton energies are derived from
            a power-law of index $-$2.1 (differential) with an exponential cutoff at 10$^{14}$eV.
           } 
   \label{fig:EGRET_Naito_w28}
  \end{figure}

The most interesting question concerns the possibly of enhanced TeV gamma-ray emission 
from regions of high ISM density (Aharonian \etal \cite{Aharonian:1}), where there is a greater
chance for the interaction of cosmic rays from the SNR. The molecular clouds 
along the northeast and northern remnant boundary have been mapped in detail by Arikawa \etal (\cite{Arikawa:1}) at the CO
J=1-0 and J=3-2 lines. The shocked (region undergone passage and compression by the SNR shock) 
component of the clouds is distributed along the SNR/clouds boundary, has a mean density of 
$\geq 10^4$ cm$^{-3}$, and mass $2\times 10^3$M$_\odot$. The other 4$\times 10^3$M$_\odot$ of the unshocked gas has a density of 
$\leq 10^3$cm$^{-3}$ and is displaced radially outward from the shocked gas by $\sim$1 arcmin.
Clearly, any TeV gamma-ray flux from $\pi^\circ$ decay would be dominated by the shocked gas regions, given that
the unshocked regions of the cloud would have a much lower density and lower energy available for CR 
production, and the mean matter density for regions excluding the molecular cloud is only $\sim$1.3 cm$^{-3}$ 
(Esposito \etal \cite{Esposito:1}). 
Arikawa \etal (\cite{Arikawa:1}) has derived the energy deposited into the shocked gas at $3\times 10^{48}$ergs, a value
consistent with the Rho \etal (\cite{Rho:1}) estimate for the total SNR energy when considering the volume 
filling factor $V$, between 
the clouds and the SNR. $V$ is simplistically estimated at $\leq 0.01$, by taking the mass and density of the 
shocked gas (given above), assuming the cloud consists of H$_2$, and a value of 10pc for the 
SNR radius. We also assume that $\sim$10\% of the available SNR energy goes into cosmic ray production, a
reasonable value for a Sedov-phase SNR, and also consistent with the measured energetics for the SNR and clouds. Thus
we adopt values of 3$\times 10^{47}$ergs for $E_{cr}$, and $10^4$ cm$^{-3}$ for $n$ in Eq.~(\ref{eq:fluxscaling}).
A working band on our flux prediction is obtained if we assume a range of values for $d$ from
1.8 to 3.3 kpc, as discussed in Sect.~\ref{sec:intro}. 
In Fig.~\ref{fig:EGRET_Naito_w28}, we compare the model predictions based on the above scalings 
to our upper limit obtained from an extended source of radius 0.25$^\circ$ encompassing the clouds
from 1994 data (the highest of our upper limits). The flux from 2EG J1801$-$2312 (essentially the
same as 3EG J1800$-$2338) and its straight extrapolation to TeV energies is also included as any $\pi^\circ$ gamma-ray flux will
be limited by the EGRET measurement. 

Our upper limit lies an order of magnitude below the straight extrapolation of the flux from EGRET  (dashed line of 
Fig~\ref{fig:EGRET_Naito_w28}), and is able to place 
some constraint on the prediction of a $\pi^\circ$ gamma-ray flux from the shocked gas region. In order to accommodate our upper 
limit however a cutoff and/or slighter steeper parent spectrum than -2.1 appears necessary. 
In assuming an exponential cutoff at 100 TeV for the parent spectrum of accelerated hadrons,  we assume that W28 follows the 'standard'
picture of particle acceleration in SNR, with particle energies limited by radiative losses, finite age of the 
shock and particle escape. When the cutoff for electrons is due only to radiative 
losses however, it can be expected that the hadron spectrum will continue (Reynolds \& Keohane \cite{Reynolds:1}). Apart from the 
long-established 'knee' at $\sim5\times10^{15}$ eV in the all-particle cosmic ray spectrum 
there is now direct experimental evidence pointing to a continuation of the proton/helium spectra up to at least $\sim 800$ TeV 
(Asakimori \etal \cite{Asakimori:1}), implying that strong cutoffs may not be required. Our result here is consistent however, 
with previous comparisons of upper limits (from other SNRs) to hadron-induced gamma-ray models 
(e.g. Buckley \etal \cite{Buckley:1}, Allen \etal \cite{Allen:1} and Prosch \etal \cite{Prosch:1}) which seem to require some 
cutoff below the knee energy/and or spectra steeper than -2.1. Further constraints may arise if electronic components are considered. 
Particularly, electronic  bremsstrahlung may dominate over the inverse Compton component due to the very high density of target matter 
in the clouds.
 

The above discussion aside, the location of 3EG J1800$-$2338 by itself, makes it's interpretation difficult in terms of simple 
CR-matter interaction, or as the result of a pulsar-powered process (proposed by Merck \etal \cite{Merck:1}). Many of the promising 
sites within W28 for gamma-ray production are now outside the 95\% error circle, leaving just the filled X-ray centre, and 
southern/western portions of the radio shell. At best, we are in a position to rule out the interpretation of the EGRET source 
as resulting totally from $\pi^\circ$ decay gamma-rays with an unlimited parent spectrum, and to place limits on
the parent spectral index/cutoff energy combination.

\section{Conclusion}
A search for TeV gamma-ray emission from the southern SNR W28 was carried out by the CANGAROO over two
observation seasons (1994 and 1995) using the atmospheric \C imaging technique.
An analysis providing a consistent gamma-ray acceptance and quality factor for extended sources 
was used. A number of sites within a search region of $\pm1^\circ$ were considered as potential point-like
and diffuse sources of TeV gamma-ray emission. 
No evidence was found for the emission of TeV gamma-rays at any of these sites which include those from 
the strongest two masers, an EGRET source, a radio pulsar (all as point sources) 
and a diffuse region containing the molecular clouds that 
appear to be interacting with the remnant.  
Our 3$\sigma$ upper limit from this diffuse region at 6.64$\times10^{-12}$ cm$^{-2}$s$^{-1}$ 
$> 1.5$ TeV, and the flux of the EGRET source 
3EG J1800$-$2338 were compared with gamma-ray flux predictions from the 
model of Naito and Takahara (\cite{Naito:1}). Under this framework, 
our upper limit rules out a straight extrapolation of the EGRET flux to TeV energies. It also 
constrains somewhat the flux expected from the shocked region of gas in the molecular cloud, placing limits
on the parent spectra for hadrons and/or cutoff energy. Our results suggest that
the EGRET source probably does not result entirely of $\pi^\circ$ gamma-rays. This fact is supported by it's location in relation
to the molecular clouds. In a later paper, we will consider electronic 
bremsstrahlung and inverse Compton scattering and discuss the broader implications of our results 
in relation to the origin of galactic cosmic rays. 

Further data on W28 will no doubt be taken with the recently completed 
CANGAROO II telescope (Yoshikoshi \etal \cite{Yoshikoshi:3}), which, at the very least will provide tighter
constraints on models of gamma-ray production for this interesting source.

 \begin{acknowledgements}
This work is supported by a Grant-in-Aid in Scientific Research from
the Japanese Ministry of Science, Sports and Culture, and also by the
Australian Research Council. GR, JH, MR \& TY acknowledge the receipt of
JSPS postdoctoral fellowships. We also thank an anonymous referee for 
valuable comments.
 \end{acknowledgements}


\begin{thebibliography}{xx}


  \bibitem[1991]{Akerlof:1} Akerlof C.W., Cawley M.F., Chantell M., et~al 1991, ApJ 377, ~L97
  \bibitem[1995]{Allen:1} Allen G.E., Berley D., Biller S., et~al 1995, ApJ 448, ~L25
  \bibitem[1978]{Altenhoff:1} Altenhoff W.J., Downes D., Pauls T., et~al 1978, A\&AS 35, ~23
  \bibitem[1994]{Aharonian:1} Aharonian F., Drury L. O'C., V\"{o}lk H.J., et~al. 1994, A\&A 285, ~645
  \bibitem[1983]{Andrews:1} Andrews M.D. Basart J.P., Lamb R.C., et~al. 1983, ApJ 266, ~684 
  \bibitem[1999]{Arikawa:1} Arikawa Y., Tatematsu K., Sekimoto Y., Takahashi T. 1999, Proc. Astron. Soc. Jap. 51, ~L7 
  \bibitem[1998]{Asakimori:1} Asakimori K., Burnett T.H., Cherry M.L., et~al 1998, ApJ. 502, ~278
  \bibitem[1999]{Baring:1} Baring M.G., Ellison  D.C., Reynolds S.P., Grenier I.A., Goret P. 1999, ApJ, 513 ~311
  \bibitem[1987]{Blandford:1} Blandford R., Eichler D. 1987, Phys. Rep. 154, ~1
  \bibitem[1998]{Buckley:1} Buckley J.H., Akerlof C.W., Carter-Lewis D.A., et~al. 1998, A\&A 329, ~639
  \bibitem[1997]{Claussen:1} Claussen M.J., Frail D.A., Goss W.M., Gaume R.A. 1997, ApJ 489, ~143
  \bibitem[1999]{Claussen:2} Claussen M.J., Goss W.M., Frail D.A., Desai K. 1999, ApJ 522, ~349 
  \bibitem[1998]{Connaugton:1} Connaugton V., Akerlof C.W., Biller S., et~al. 1998, Astropart. Phys. 8, ~179 
  \bibitem[1994]{Drury:1} Drury L. O'C, Aharonian F., V\"{o}lk H.J. 1994, A\&A 287, ~959
  \bibitem[1996]{Esposito:1} Esposito J.A., Hunter S.D., Kanbach G., Sreekumar P. 1996, ApJ 461, ~820
  \bibitem[1993]{Frail:1} Frail D.A., Kulkarni S.R., Vasisht G. 1993, Nat. 365, ~136
  \bibitem[1998]{Gaisser:1} Gaisser T.K., Protheroe R.J., Stanev T. 1998, A\&A. 492, ~219
  \bibitem[1999]{Goret:1} Goret P., Gouiffes C., Nuss E., et~al. 1999, Proc. 26th Int. Cosmic Ray Conf. (Salt Lake City)
                                              3, ~496
  \bibitem[1976]{Goudis:1} Goudis C., 1976, Ap\&SS 40, ~91
  \bibitem[1998]{Green:1} Green D.A., 1998, `A Catalogue of Galactic Supernova Remnants (1998 September version)', Mullard Radio Astronomy Observatory, Cambridge, United Kingdom (available on the World-Wide-Web at {\tt http://www.mrao.cam.ac.uk/surveys/snrs/}). 
  \bibitem[1993]{Hara:1} Hara T., Kifune T., Matsubara Y., et~al 1993, Nucl. Inst. Meth. 300, ~A332
  \bibitem[1999]{Hartman:1} Hartman R.C., Bertsch D.L., Bloom S.D., et~al 1999, ApJS 123, ~79
  \bibitem[1997]{Hess:1} Hess M. 1997, 
                Proc. 25th Int. Cosmic Ray Conf. (Durban) 3, ~229
  \bibitem[1985]{Hillas:1} Hillas A.M. 1985, 
                Proc. 19th Int. Cosmic Ray Conf. (La Jolla) 3, ~445
  \bibitem[1991]{Hillas:2} Hillas A.M., West M. 1991, 
                Proc. 22nd Int. Cosmic Ray Conf. (Dublin) 1, ~472   
  \bibitem[1995]{Hillas:3} Hillas A.M. 1995,
                Proc. 24th Int. Cosmic Ray Conf. (Rome) 1, ~270 
  \bibitem[1993]{Kaspi:1} Kaspi V.M., Lyne A.G., Manchester R.N., et~al 1993, ApJ 409, ~L57 
  \bibitem[1993]{Kifune:1} Kifune T. 1993, 
                        Proc. 23rd Int. Cosmic Ray Conf. (Calgary) 1, ~444 
  \bibitem[1997]{Lamb:1} Lamb R.C., Macomb D.J. 1997, ApJ 488, ~872
  \bibitem[1999]{Lessard:1} Lessard R.D., Bond I.H., Boyle P.J., et~al 1999, Proc. 26th Int. Cosmic Ray Conf. (Salt Lake City)
                                              3, ~488
  \bibitem[1983]{Li:1} Li T., Ma Y. 1983, ApJ 272, ~317
  \bibitem[1991]{Long:1} Long K.S., Blair W.P., White R.L., Matsui Y., 1991, ApJ 373, ~567
  \bibitem [1981]{Lozinskaya:1} Lozinskaya T.A. 1981 Sov Astron Lett 7, ~17 
  \bibitem[1996]{Mastichiadis:1} Mastichiadis A., de Jager O.C. 1996, A\&A 311, ~L5
  \bibitem[1996]{Merck:1} Merck M., Bertsch D.L., Dingus B.L., et~al 1996, A\&AS 120, ~465 
  \bibitem[1995]{Mori:1} Mori M. 1995, 
                       Proc. 24th Int. Cosmic Ray Conf. (Rome) 4, ~487 
  \bibitem[2000]{Muraishi:1} Muraishi H., Tanimori T., Yanagita S., et~al 2000, A\&A {\em in press}
  \bibitem[1994]{Naito:1} Naito T., Takahara F. 1994, J Phys G: Nucl Part Phys 20, ~477
  \bibitem[1996]{Prosch:1} Prosch C., Feigl E., Plaga, R., et~al 1996, A\&A 314, ~275
  \bibitem[1999]{Puhlhofer:1} P\"{u}hlhofer G., V\"{o}lk H., Wiedner C.A., et~al 1999, 
                 Proc. 26th Int. Cosmic Ray Conf. (Salt Lake City) 3, ~492   
  \bibitem[1999]{Reynolds:1} Reynolds S.P., Keohane J.W. 1999, ApJ 525, ~368 
  \bibitem[1998]{Rho:1} Rho J.H., Petre R. 1998, ApJ 503, ~L167 
  \bibitem[1996]{Rho:2} Rho J.H., Petre R., Pisarski R., Jones L.R. 1996, MPE Report 263, ~273
  \bibitem[1998]{Roberts:1} Roberts M.D., McGee P., Dazeley S.A., et~al 1998, A\&A 337, ~25
  \bibitem[1999]{Romero:1} Romero G.E., Benaglia P. and Torres D.F. 1999, A\&A 348, ~868
  \bibitem[1999]{Rowell:1} Rowell G.P. Dazeley S.A., Edwards P.G., et~al 1999, Astropart. Phys. 11, ~217 
  \bibitem[1998]{Tanimori:1} Tanimori T., Hayami Y., Kamei S. et~al 1998, ApJ 492, ~L33 
  \bibitem[1996]{Thompson:1} Thompson D.J., Bertsch D.L., Dingus B.L., et~al 1996, ApJS 107, ~227
  \bibitem[1998]{Tomida:1} Tomida H. 1998, Genshikaku Kenkyu 42, ~123, {\em in Japanese}
  \bibitem[1981]{Wootten:1} Wootten A. 1981, ApJ 245, ~105
  \bibitem[1996]{Yoshikoshi:2} Yoshikoshi T. 1996, PhD. thesis, Tokyo Institute of Technology.
  \bibitem[1997]{Yoshikoshi:1} Yoshikoshi T., Kifune T., Dazeley S.A., et~al 1997, ApJ 487, ~L65
  \bibitem[1999]{Yoshikoshi:3} Yoshikoshi T., Dazeley S.A., Gunji S., et~al 1999, Astropart. Phys. 11, ~267
  \bibitem[1998]{Zhang:1} Zhang I., Cheng K.S. 1998, A\&A 335, ~234
 \end{thebibliography}
\end{document}